\documentclass[reprint]{revtex4-1}

\usepackage{bbold} 
\usepackage{graphicx}
\usepackage{verbatim}
\usepackage{mathtools}
\usepackage{url}
\usepackage{amsthm} 
\usepackage{hyperref} 
\usepackage{xfrac}
\usepackage{nicefrac}

\usepackage{amssymb}

\begin{document}

\title{Predictive Maxwell's Demons}

\author{Nathaniel Rupprecht, Dervis Can Vural}
\affiliation{University of Notre Dame, South Bend, IN}

\date{\today}

\begin{abstract}
Here we study the operation efficiency of a finite-size finite-response-time Maxwell's demon, who can make future predictions. We compare the heat and mass transport rate of predictive demons to non-predictive ones and find that predictive demons can achieve higher mass and heat transport rates over longer periods of time. We determine how the demon performance varies with response time, future sight, and the density of the gasses on which they operate.
\end{abstract}

\maketitle

{\bf Introduction.} A Maxwell's demon is a device that makes microscopic measurements and sort molecules according to their energy or momentum, thereby reducing entropy \cite{maxwell1891theory}. It has served as a pivotal thought experiment to establish the close relationship between thermodynamics, measurement, and information processing \cite{landauer,thomson1874kinetic,smoluchowski1927experimentell,szilard1929entropieverminderung,brillouin1951maxwell}.

Maxwell's original thought experiment has recently been expanded to include feedback control \cite{cao2004feedback,seifert2012stochastic} and universal computation \cite{zurek1999algorithmic,caves1990quantitative,caves1990comment}. Some authors describe the computational process within the demon as manipulations of a tape bits \cite{mandal2012work,mandal2013maxwell,barato2013autonomous,hosoya2015operational} or qubits \cite{deffner2013information}, while others consider demons that manipulate microstates by fully mechanical means \cite{gordon1983maxwell,skordos1992maxwell,tu2008efficiency}. Models of non-ideal demons \cite{mandal2012work,mandal2013maxwell} account for the thermal equilibriation of the demon with the system, the demon's finite mass \cite{tu2008efficiency}, or finite size and response time \cite{rupprecht2019maxwells}. Recently experimental constructions have also become feasible \cite{strasberg2013thermodynamics,roldan2014universal,cottet2017observing,thorn2008experimental,bannerman2009single,koski2015chip,vidrighin2016photonic,camati2016experimental,cottet2017observing,chida2017power,schaller2011probing,esposito2012stochastic,schaller2018electronic}.



Previously, we studied how the finite size and response time constrains a demon's heat and mass transport rate \cite{rupprecht2019maxwells}. However, our model demon only used local information, and did not take into consideration future arrivals. 

Knowing the future would fundamentally change the actions of a demon. A predictive demon would reject a desirable particle, if it knew that more undesirable particles will arrive before it can close the gate.

In the present study, we establish heat and mass transport limitations of a predictive Maxwell's demon with finite size and response time, and compare these limitations to that of a non-predictive demon. This way, we aim to begin exploring the thermodynamic consequences of prediction making.



{\bf Non-Predictive Demons.} Our definition of the non-predictive demon comes from a previous work \cite{rupprecht2019maxwells}, where we studied demons with finite size $A$ and response time $\tau$. We start with a brief overview of the demon model, which will be identical for both the predictive and non-predictive cases.

Our system consists of left and right subsystems of ideal gas with volumes \(V_l,\,V_r\), energies \(E_l,\,E_r\), numbers \(N_l,\,N_r\) of particles of mass \(m\). We theoretically consider \(d=1,\,2,\) or \(3\) dimensional systems, but only simulate \(d=2\). The subsystems are separated by a gate of area \(A\) (\(A\equiv1\) for a 1D system, and to be the length of the gate for a two dimensional system) controlled by the demon. We assume that the subsystems are large enough that each subsystem acts as a self averaging canonical distribution. We take \(N_s,\,E_s \to \infty\) with fixed \(\rho_s \equiv N_s/V_s\) and \(\bar{E}_s \equiv E_s/V_s\),
where the temperature, \(1/\beta_s\), and energy per particle \(\bar{E}_s =E_s/N_s\) are related by \(\beta_s E_s = N_s  d/2 \equiv \beta_s N_s \bar{E}_s\), as required by the equipartition theorem.

\begin{figure}
    \centering
    \includegraphics[width=0.48\textwidth]{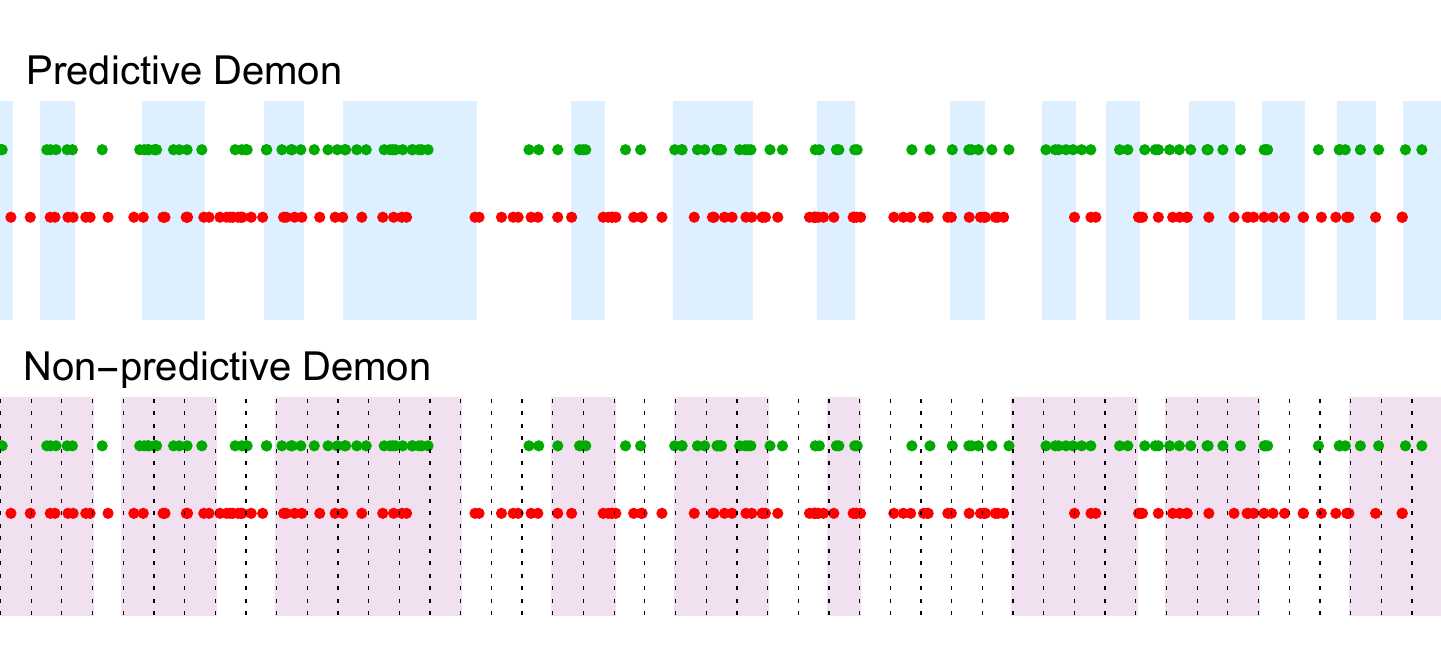}
    \caption{The operation of a predictive (top) and a non-predictive demon (bottom). Shaded regions indicate times when the gate was open. Green and red dots indicate the arrival of particles from left and right. While the non-predictive demon decides whether to open or close its gate based on particle arrivals within one \(\tau\), the predictive demon plans ahead.}
    \label{Fig:Comparison}
\end{figure}

We analyze two types of demons. (1) An \emph{energy} demon opens the gate whenever the net flow of energy from left to right would be positive. (2) A \emph{number} demon opens the gate whenever the net particle transfer from left to right is positive.

The probability that \(n\) particles are incident on the gate during a length of time \(\tau\) follows a Poisson distribution with a Poisson parameter \cite{rupprecht2019maxwells},
\begin{align}
    \kappa_s &= \frac{\rho_s \tau A}{\sqrt{2 \pi \beta_s m}} = \rho_s\tau A \sqrt{\frac{\bar{E}_s}{d\,\pi\,m}} \equiv \nu_s \tau.
    \label{kappa}
\end{align}
where $s=l,r$ for the left and right subsystem.
The rates \(\nu_l,\,\nu_r\) are crucial in characterizing the demons' performance. For example, the average energy and number currents for non-predictive demons with small response time \(\tau\) are \cite{rupprecht2019maxwells}
\begin{align*}
    \dot{E}^{(d)}_\tau &= [(d+1)/(2\beta_l)]\nu_l e^{-\nu_r \tau} + \mathcal{O}\left(\tau\,e^{-(\nu_l+\nu_r)\tau}\right)
    \\
    \dot{N}^{(d)}_\tau &=\nu_l e^{-\nu_r \tau} + \mathcal{O}\left(\tau\,e^{-(\nu_l+\nu_r)\tau}\right).
\end{align*}

As mentioned before, a non-predictive demon simply ``seizes the day'' by making decisions only according to the mass or energy flux during a present time interval \(\tau\). We will see that a predictive demon on the other hand will be able to forgo short term success for higher average transmission rates on the long run.

{\bf Predictive demons.} Given a sequence of times at which individual particles hit the gate area from the left or right, the predictive demon must determine the optimal sequence of gate openings and closings, subject to the response time constraint, i.e. the gate cannot change its open/close state faster than \(\tau\).

We will suppose that particle arrivals to the gate area are independent events, and we will assign a ``score'' to each arrival according to a suitable characteristic of the particle, which the demon aims to maximize. For a number demon, the score will be $1$ and $-1$ for left and right arrivals. For an energy demon, the score associated with an event is the energy of the particle, again with a sign that depends on its direction.


As a benchmark to evaluate how well the predictive demon is doing, we will compare it to the non-predictive demon. We do allow the non-predictive demon to decide what offset \(0 \le t_0 < \tau\) its time division should have. The demon will then bin events into the bins \([k\tau-t_0, (k+1)\tau-t_0]\), \(k\ge0\), and decide for each bin whether it is better for the gate to be open or closed. When the total time of the simulation is long, the offset does not effect the average score.  Fig. \ref{Fig:Comparison}, which shows how the two types of demons treat the same set of events.

{\bf Simulating a predictive demon} We created a program that determines the schedule of gate openings and closings for predictive and non-predictive demons, for a given set of random left and right particle arrivals with Boltzmann-distributed energies \footnote{See \url{https://github.com/nrupprecht/Scheduling-Process} for our implementation.}. To implement the scheduling, the predictive demon divides time into very small \emph{microbins}, and then converts scheduling into a discrete problem by placing particle arrival events into these microbins. We call the number of microbins per \(\tau\) the \emph{resolution}, $g$, of the simulation. We take the large \(g\) limit to approximate the continuous time problem. We have found that demon performance does not change much beyond \(g=50\) (see \ref{Appendix:resolution}).

Using our program, we numerically study how the efficiency of predictive demons change and compares to the non-predictive demons as we change subsystem temperatures, number densities and demon response time. We do so by generating a realization of particle arrival times and scores, and pass this to the scheduler program. The energy of a particle given that it hits the gate area follows a Gamma distribution \cite{rupprecht2019maxwells},
\(\hat{E}_{l,r}\vert_{\text{hit}} \sim \hat{\Gamma}\left((d+1)/2, \beta_{l,r}\right)\)
which we use to assign energy scores. 

The \emph{performance} of the number and energy demons is quantified by the average mass and heat transfer (average score per unit time) they can achieve.

\begin{figure}
    \centering
    \includegraphics[width=0.5\textwidth]{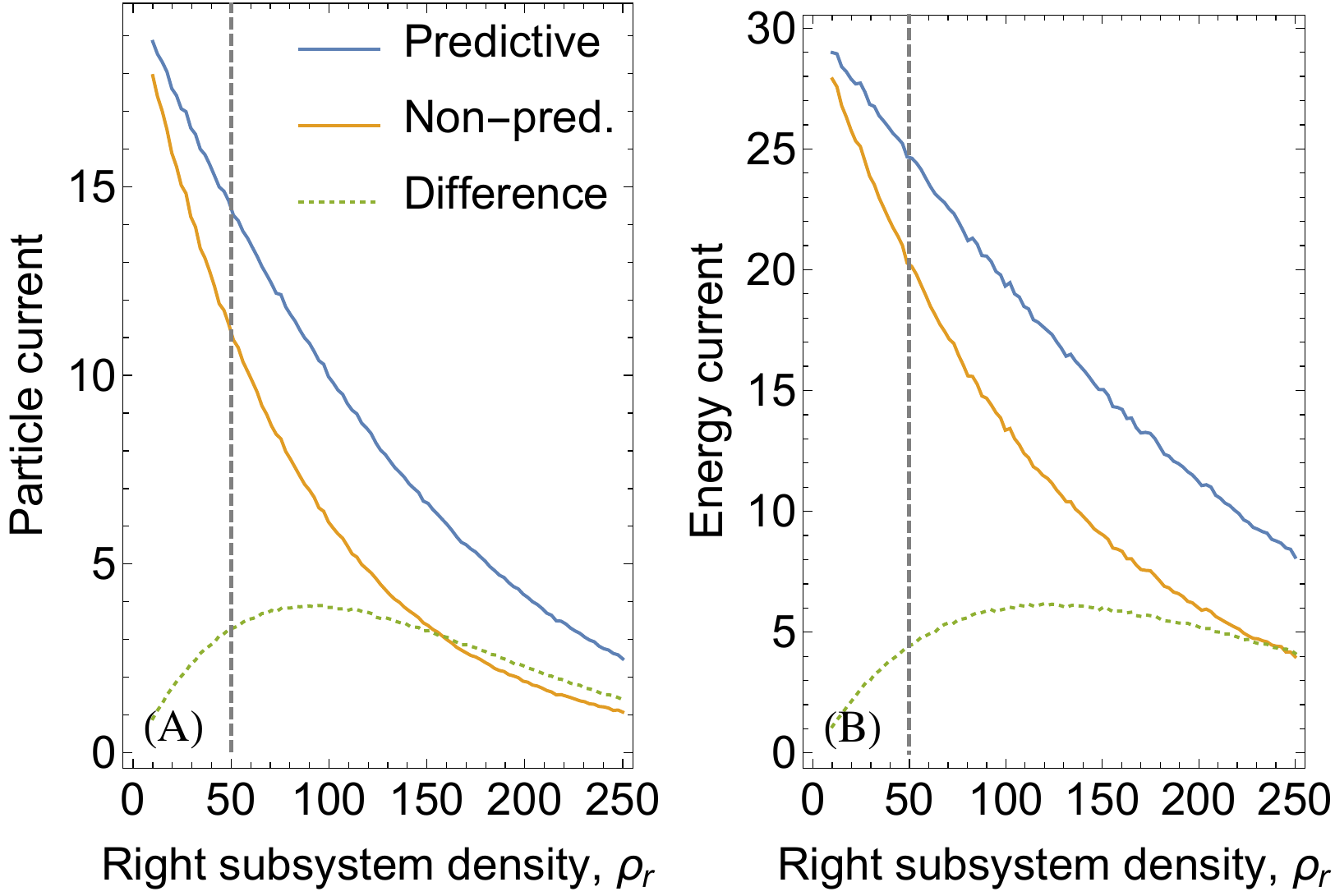}
    \caption{\textbf{Performance versus density.} 
    The performance of number demons (panel A) and energy demons (panel B) for predictive and non-predictive (orange) cases are plotted as a function of right subsystem density, while the left subsystem density is kept constant (vertical dashed line), averaged over 500 particle arrival sequences.}
    \label{Fig:VarySystem}
\end{figure}

{\bf Thermodynamic consequences of prediction making.} We start by studying the basic properties of a who can see infinitely far into the future (\(t_s \to \infty\)) and determine how it compares to a non-predictive demon.

First, we show how the demons' performance varies with the system parameters (Fig. \ref{Fig:VarySystem}). 
The number density of the left subsystem is fixed at \(\rho_l = 50\). We find that for both number and energy demons, the predictive demon has the maximal advantage over the non-predictive demon when the right subsystem is denser, but not too much denser, than the left subsystem.

\begin{figure}
    \centering
    \includegraphics[width=0.5\textwidth]{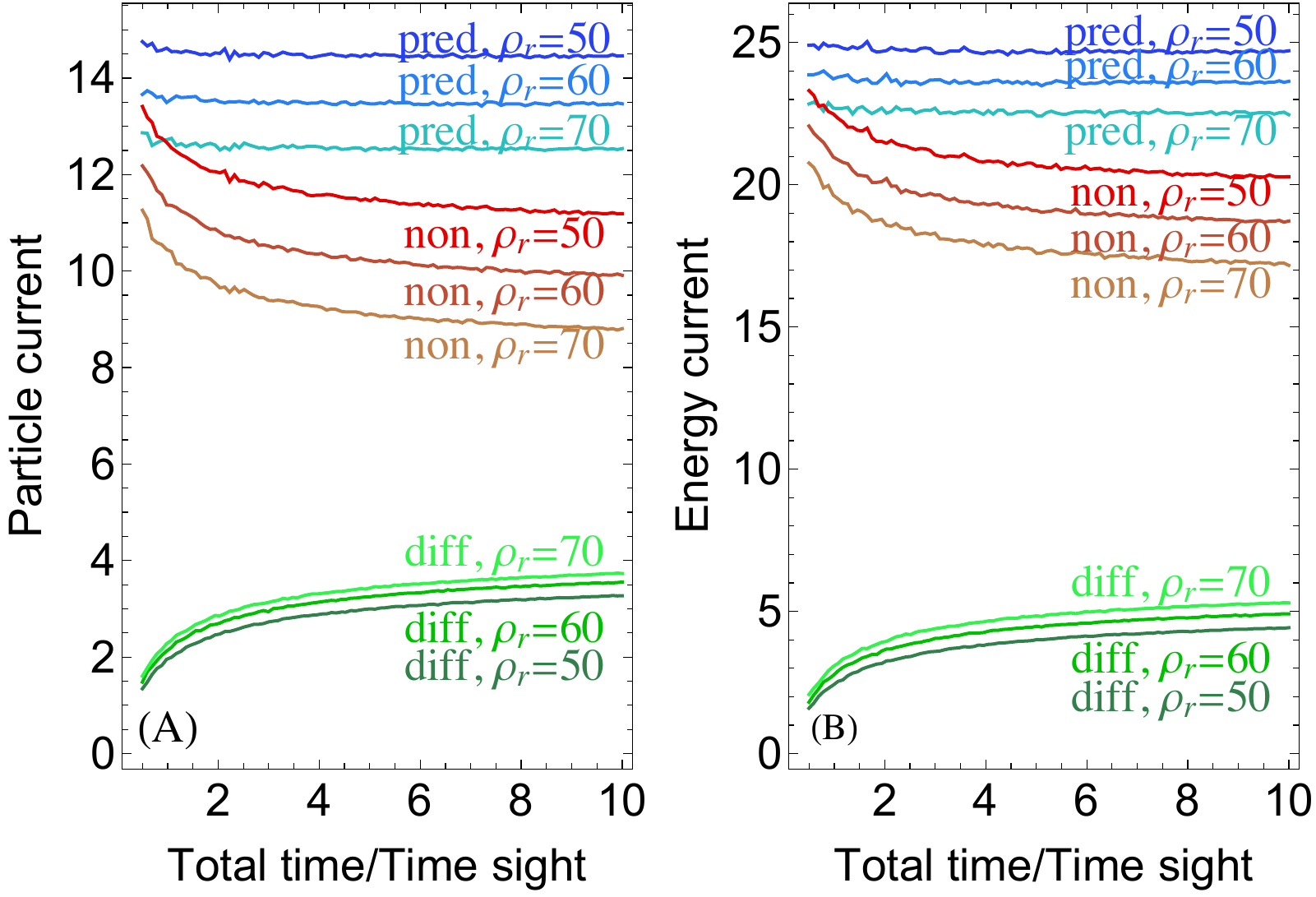}
    \caption{\textbf{Performance vs. total time.} Performance was average over 5000 random sets of events. (A) Predictive (blue) and non-predictive (red) demons and their difference (green) is plotted for number (panel A) and energy demons (panel B).}
    \label{Fig:VaryTime}
\end{figure}

Secondly, we study how the demons' performance changes as the total time they must schedule over, \(T\), increases (again, \(t_s\to\infty\)). Intuitively, as \(T\) increases, both demons should do worse. To see why, suppose that the average score on an interval of time of length \(t\) is \(S_t\). Then the average performance on two separate intervals of length \(t\) is just \(2\,S_t\). However, if these time intervals are not separated, but contiguous, then the ``boundary conditions'' of the best schedules on each time interval will in general not align. 

The demon performance as a function of total time $T$ is shown for several different right subsystem parameters in Fig. \ref{Fig:VaryTime}. The heat and mass currents for the predictive demons quickly asymptote (blue curves), whereas that for the non-predictive demons asymptote far later (red curves). A predictive demon's advantage over the non-predictive demon continues to grow long after the change in its performance has stabilized. We also see that while the performance of both demons drop when the density of the right subsystem increases, the difference in performance increases for both the number and energy demons.

If we had restricted the non-predictive demon so that its binning offset, \(t_0\), was zero, then changing the length of time it must operate over would not change the performance - its performance would always be at a minimum. Clearly, since the performance of the non-predictive demon decreases by a large amount as \(t_\text{sched}\) (see \ref{Fig:VaryTime}), it is a big advantage for the non-predictive demon to be able to determine its time offset, at least for short times. Indeed, in the limiting case where \(t_\text{sched} \le \tau\), both types of demons operate in exactly the same way. 

\begin{figure}
    \centering
    \includegraphics[width=0.5\textwidth]{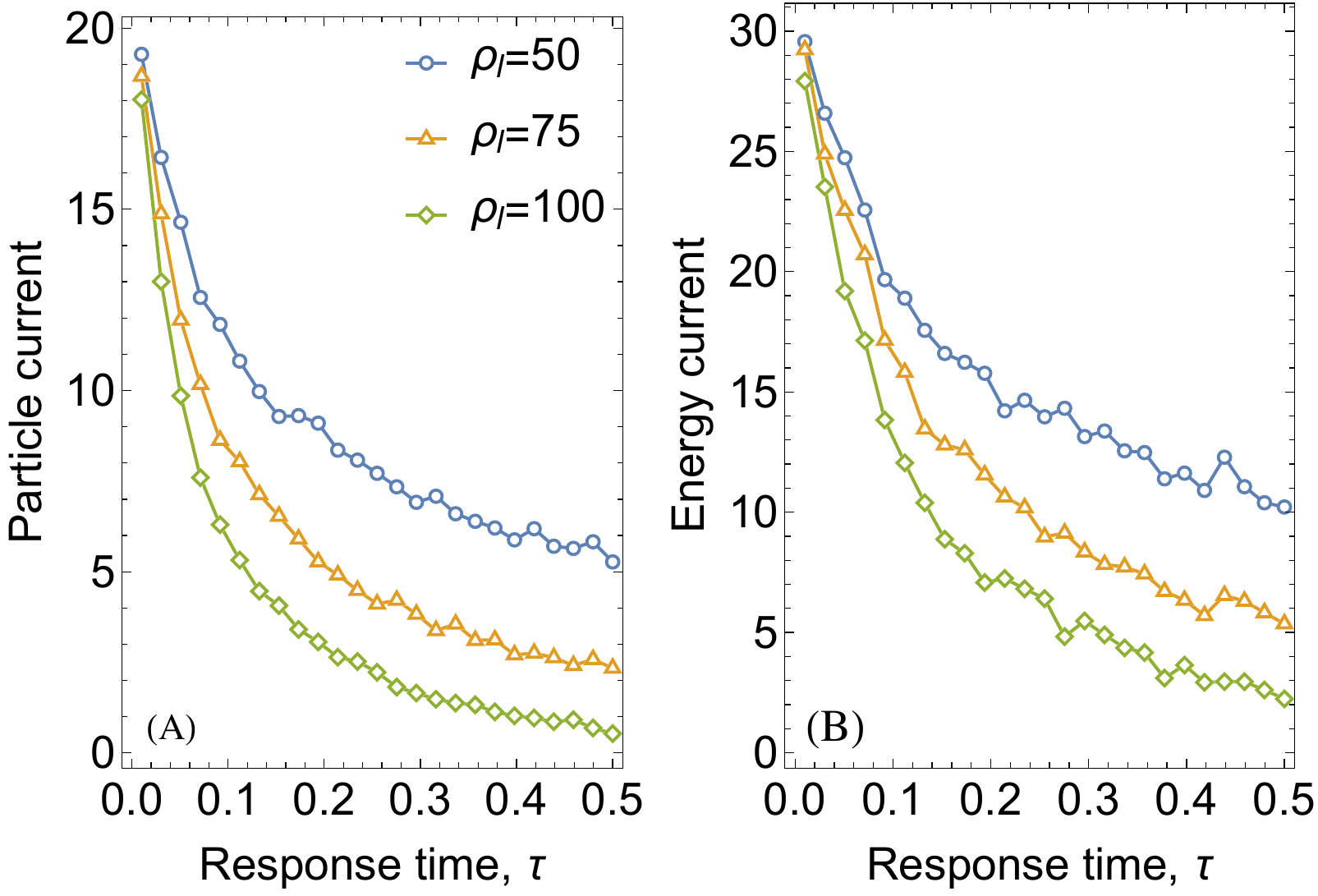}
    \caption{\textbf{Number and energy currents generated by predictive demons, as a function of response time.} For all runs, the total time was \(T=10\), and the time sight was \(t_s=1\). (A) The performance of a number demon as response time changes. (B) The performance of the energy demon as response time changes.}
    \label{Fig:RateVsTau}
\end{figure}

Third, we move on to demons who have finite future sight $t_s$, operating for a long time \(T \gg t_s\). Simulating this demon is more complicated than simply scheduling gate openings and closings over a fixed amount of time that contains all the events. Since the demon only has knowledge of the events for a small part of the total time, it must revise its belief of what the best course of action is whenever it becomes aware of new information. The demon also must also contend with the fact that the gate will most likely have to remain open or closed for some period of time beyond \(t_s\). 
If the subsystems parameters are such that it is very likely that the net flow of particles will be negative during this time, then the demon should consider choosing a schedule that results in a lower score during the interval \([0, t_s]\), but does not require the gate to be open as long during the period after \(t_s\). 

Since part of the schedule extends into a time the demon cannot see, the score is now a random variable that can be decomposed into a sum of the (deterministic) score \(X_d\), of the schedule in the interval \([0, t_s]\), and the (probabilistic) score \(\hat{X}_r\) of the part of the schedule beyond \(t_s\).

When calculating the projected score for the schedule, we take the expectation value, \(X_d + \langle \hat{X}_r \rangle\), to determine which schedule is best. We have observed that not taking the random part of the scheduling score into account results in very low or even negative demon performance as \(\tau\) increases.

\begin{figure}
    \centering
    \includegraphics[width=0.5\textwidth]{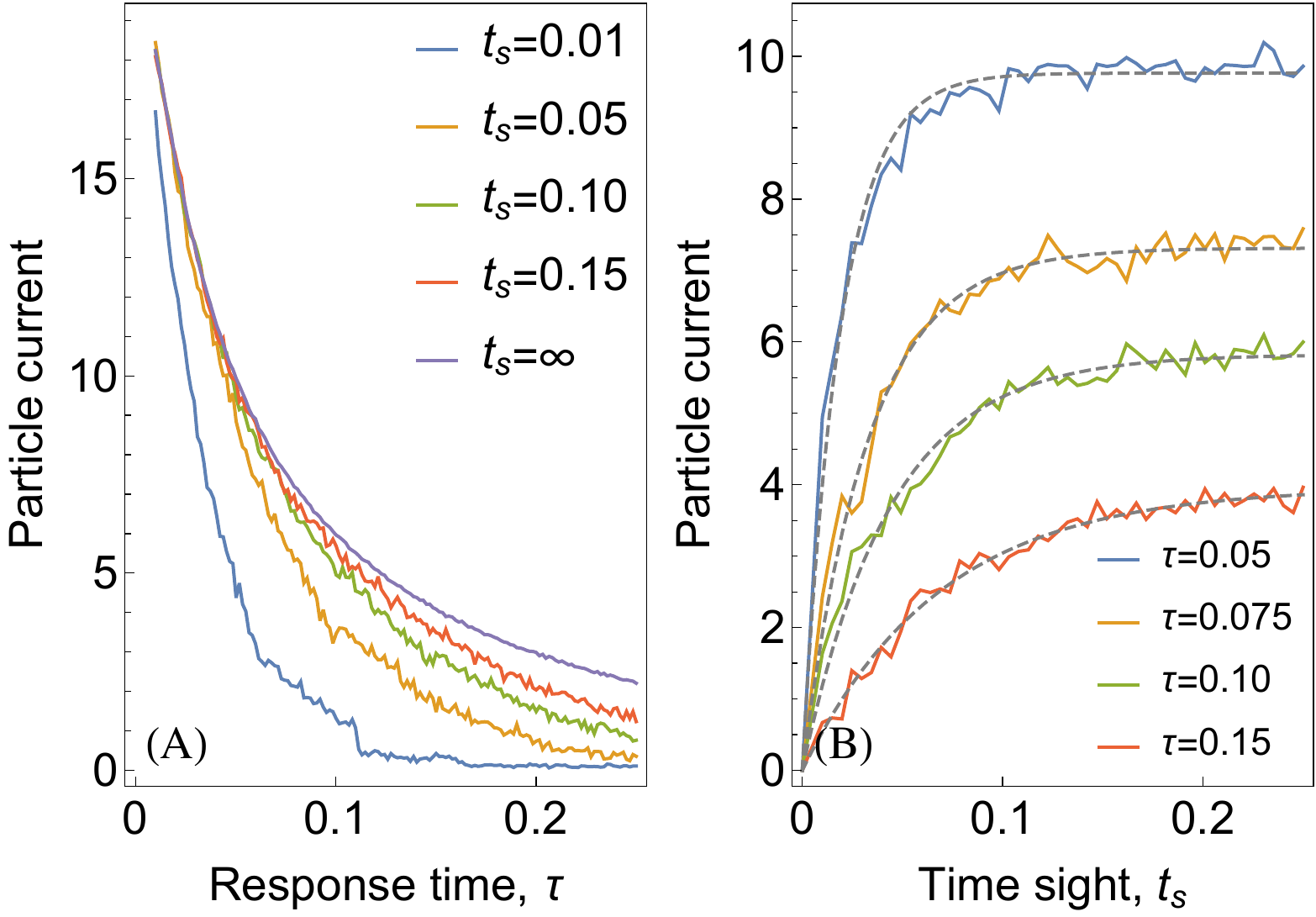}
    \caption{\textbf{Number demon performance, varying response time and time sight.} The systems had parameters \(\rho_l = 50,\,\rho_r=100\). (A) Number rate vs. response time for several fixed time sights. An averaging of 500 runs with \(t=1\) second was used. (B) Number rate vs. time sight for several fixed response times. The best fitting functions of the form \(\dot{N}_\tau(t_s) = a_\tau(1-e^{-\lambda_\tau\,t_s})\) for each \(\tau\) are shown in gray.}
    \label{Fig:RatesVsTimes}
\end{figure}

The performance of demons with fixed time sight (\(t_s = 1.0\)) and varying response times is shown in Fig.\ref{Fig:RateVsTau}, and is similar to that in \cite{rupprecht2019maxwells}, where for \(\tau \approx 0\), the rates decrease exponentially, but as \(\tau\) increases, additional terms of higher orders slow down the decrease. We also check what happens when the time sight is kept fixed and the response time varies, \ref{Fig:RatesVsTimes}(a), and when the response time is fixed and the time sight varies, \ref{Fig:RatesVsTimes}(b). The qualitative behavior of predictive energy demons is similar (see Fig. \ref{Fig:RateVsTau}).

In Fig. \ref{Fig:RatesVsTimes}(a), we see that having larger time sight is strictly better. 
Fig. \ref{Fig:RatesVsTimes}(b) shows demons with fixed response time as \(t_s\) is varied. Of course, faster response time is strictly better.  
While most of the curves that we found in \cite{rupprecht2019maxwells} and here, including \(\dot{N}_\tau(\infty)\), are ``smoothly decreasing'', they cannot be fit to simple exponential or rational functions. Interestingly, while the infinite time sight rates, \(\dot{N}_\tau(\infty)\), are non-trivial functions, the correction for finite time sight appears to be quite simple,
\begin{align}
    \dot{N}_\tau(t_s) = (1-e^{-\lambda_\tau\,t_s}) \dot{N}_\tau(\infty).
    \label{FiniteTimeSight}
\end{align}
Despite our best efforts, we were not able to derive this elementary relationship.


{\bf Analytical considerations.} Assuming that a demon has \(t_s = \infty\) and a total scheduling time \(T\), for any response time \(\tau\), microbin time \(dt = \tau/g\), and a random variable \(\hat{v}\) that encodes the distribution and score of events, we can define the discrete Scheduling process, \(\hat{\mathbf{S}}_t = \hat{\mathbf{S}}_t[\tau,\,dt,\,\hat{v}, T] \equiv \{\hat{S}_t,\,\hat{S}_t^*\}\), which is the best score possible during the remaining time in the scheduling interval given that the door was open for at least the previous \(\tau\) or closed for at least the previous \(\tau\) (\(\hat{S}_t\) and \(\hat{S}_t^*\), respectively). The scheduling process is defined by the set of equations
\begin{align}
    \hat{S}_t &= \text{max}\left(\hat{S}_{t+dt} + \hat{v}_t(dt),\, \hat{S}_{t+\tau}^* \right)
    \label{SchedulingProcessOpen}
    \\
    \hat{S}^*_t &= \text{max}\left(\hat{S}_{t+\tau} + \hat{v}_t(\tau),\, \hat{S}_{t+dt}^* \right)
    \label{SchedulingProcessClosed}
\end{align}
where \(\hat{v}_t(\Delta t)\) is the event score between \(t\) and \(t+\Delta t\).

If we could solve for the average growth rate,  \(\langle \hat{S}_T / T \rangle\), of these equations as \(T \to \infty\), we would obtain the performance of a predictive demon with infinite \(t_s\); and since we know from (\ref{FiniteTimeSight}) that the finite sight demon is related to the infinite sight demon in a simple way, we would essentially have a full solution to the behavior of cyclically operating predictive demons (though we would have to find a way to obtain \(\lambda_\tau\)). We could also take \(dt \to 0\) to obtain the continuous time limit. 


Equations like (\ref{SchedulingProcessOpen}, \ref{SchedulingProcessClosed}) are called \emph{Bellman equations}, used extensively in mathematical optimization and dynamic programming \cite{barron1989bellman,kirk2004optimal}. These equations break the problem down into recursively computable pieces and are solved backwards in time. 
Here, we set \(\hat{S}_t = 0\) and \(\hat{S}^*_t = 0\) for \(t>T\), and solve (\ref{SchedulingProcessOpen}, \ref{SchedulingProcessClosed}) backwards in time. The two component processes represent the best possible score from time \(t\) onward \emph{given} that the door was previously open (\(\hat{S}_t\)), or given that it was previously closed (\(\hat{S}^*_t\)). At each time point, we have already computed the values of \(\hat{S}_{t^\prime},\,\hat{S}^*_{t^\prime}\) for \(t^\prime>t\), and know the random variable  \(\hat{v}\). We simply have to decide whether it would be better to have an open door or closed door, and the maximum score is just the larger of these potential scores. 

Eq. (\ref{SchedulingProcessOpen}) says that the if the door was open, the demon can keep it open for the next \(dt\) and the re-evaluate if the door should be opened or closed (resulting in a score of \(\hat{S}_{t+dt}+\hat{v}_t(dt)\)), or the demon can shut the door, which must remain shut until \(t+\tau\), and then decide what the best course of action is at that point in time, given that the door had been shut (resulting in a score of \(\hat{S}_{t+\tau}^*\)). The best score at this point in time is simply the max of these two scores. The reasoning behind (\ref{SchedulingProcessClosed}) is similar. Our simulation simply automatically solves these equations for given realizations of the events. 
To ensure that our algorithm operates correctly, we have run brute force searches over all possible sequences of door openings/closings for some feasible times and microbin sizes. If the number of total microbins becomes too large, the brute force method quickly becomes intractable. For the realizations that we checked, we see that the Bellman algorithm does correctly compute the best possible schedule and score for the demon.

While there is literature on the asymptotic behavior of stochastic algorithms \cite{pelletier1998almost}, on differential equations containing max/min terms \cite{liu2015multiple}, and on stochastic Bellman equations \cite{evans1979optimal,turhan2011deterministic}, it seems like an analytical solution to (\ref{SchedulingProcessOpen}, \ref{SchedulingProcessClosed}) would be very hard to come by, especially since the problem involves coupled stochastic processes, and we have not yet been able to solve for the average behavior of \(\hat{S}_t\) or \(\hat{S}_t^*\).

{\bf Discussion.} We have developed an optimal protocol for \emph{predictive} Maxwell's demons, determined their heat and mass transfer rates, and compared these to the performance of their non-predictive counterparts. Knowing the future greatly enhances heat and mass transport performance. 

In closing, we should emphasize that (1) The limitations on heat/mass currents reported here stem from the finite response time of the demons (which may be due to the inertia of the gate or time required to measure and process information). A demon who could measure and haul particles at infinite velocity could of course achieve infinitely large entropy reduction rates, whether predictive or non-predictive. The fact that fundamental physics prohibits infinitely-fast measurement and gate motion suggests to us that information driven heat and mass transfer (and thus entropy reduction rate) is bounded by fundamental physics. (2) Being able to predict the future does not, of course, provide additional free negative entropy. The total entropy that can be pumped out of the system is set by the number of erasures the demon must carry out during measurement and information processing, as set by Landauer's principle \cite{landauer}. Rather, prediction making improves the \emph{rate} of entropy reduction, and heat/ mass transport.


\clearpage

\onecolumngrid

\setcounter{section}{0}
\renewcommand{\thesection}{Appendix \Alph{section}}

\section{Time resolution and demon performance} \label{Appendix:resolution}
In this section, we show how the resolution of the demon's scheduling procedure effects its performance. Recall that we divide time into ``microbins'' with a certain resolution \(g\), which is the number of microbins in a response time, \(\tau\) (so the length of a microbin is \(\tau_\text{mb} = \tau/g\)). In principle, if a real demon was trying to schedule its sequence of gate openings/closing with an algorithm similar to ours, there would be a tradeoff between the amount of time and computational resources necessary to do the scheduling and the resolution the demon would use, but in the paper, we simply use a ``large'' resolution to approximate the continuous limit.

\begin{figure}
    \centering
    \includegraphics[width=0.5\textwidth]{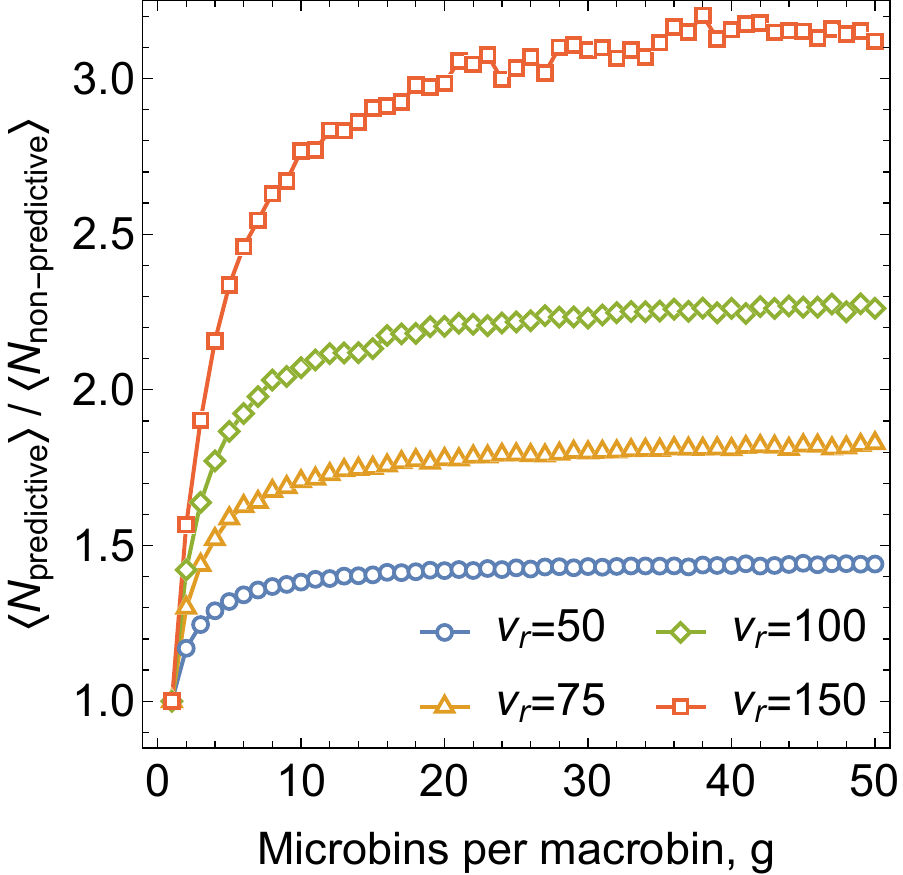}
    \caption{Convergence of a number demon's performance as the number of microbins increases, normalized by the performance of a non-predictive number demon. As \(g\to\infty\), the discrete algorithm for the demon better and better approximates the continuous demon model. Clearly, the discrete algorithm with \(g=50\) is a good approximation to the continuous case, as the score ratio has stabilized by that point.}
    \label{Fig:Microbins}
\end{figure}

As visible in Fig. \ref{Fig:Microbins}, demon score increases rapidly as the resolution increases, especially for the case where the subsystems have similar parameters.

When there is only one microbin, predictive and non-predictive demons act exactly the same, hence the ratio of scores being one. As the number of microbins increases, the predictive demon has more freedom to adjust exactly when it opens and closes its gate. Because of this, the performance of the demon rapidly increases with granularity, before asymptotically approaching the score of an ideal, continuous predictive demon.

\bibliography{bibliography.bib}

\end{document}